\documentclass[10pt]{iopart}
\usepackage{graphics}
\begin{document}
\title{Robust Bayesian detection of unmodelled bursts}
\author{Antony C Searle$^1$, Patrick J Sutton$^2$, Massimo Tinto$^3$ and Graham Woan$^4$}
\address{$^1$ Centre for Gravitational Physics, Department of Physics, Faculty of Science, The Australian National University, Canberra ACT 0200 Australia}
\address{$^2$ LIGO - California Institute of Technology, Pasadena, CA 91125}
\address{$^2$ School of Physics and Astronomy, Cardiff University, Cardiff CF24 3AA, United Kingdom}
\address{$^3$ Jet Propulsion Laboratory, California Institute of Technology, Pasadena, CA 91109}
\address{$^4$ Department of Physics and Astronomy, Kelvin Building, Glasgow G12 8QQ, United Kingdom}

\ead{$^1$\mailto{antony.searle@anu.edu.au}, $^2$\mailto{psutton@ligo.caltech.edu}, $^3$\mailto{mtinto@mail1.jpl.nasa.gov}, $^4$\mailto{graham@astro.gla.ac.uk}}
\newcommand{\gursel}{G\"{u}rsel}
\begin{abstract}
A Bayesian treatment of the problem of detecting an unmodelled gravitational wave burst with a global network of gravitational wave observatories reveals that several previously proposed statistics have implicit biases that render them sub-optimal for realistic signal populations.
%
%
%
\end{abstract}
\pacs{04.80.Nn,07.05.Kf,95.55.Ym}
\submitto{\CQG}
\maketitle
\section{Introduction}
\gursel\ and Tinto \cite{GuTi:89} first noted that a system of \emph{three} ground-based interferometers sensitive to \emph{two} incoming gravitational wave polarisations over-determines the signal, permitting a consistency check in the absence of knowledge of the waveform.  Their statistic was based on the failure to reject the null hypothesis that the observations were consistent with a gravitational wave from a postulated direction.  Subsequent authors generalised the \gursel-Tinto statistic to more than three interferometers and supplemented it with other consistency tests \cite{Ch_etal:06}, and proposed various alternatives related to it \cite{KlMoRaMi:05,KlMoRaMi:06,MoRaKlMi:06,Ra:06}.  With the partial exception of \cite{AnBrCrFl:01}, the use of Bayesian inference in the literature on coherent detection of unmodelled bursts has been confined to the justification of individual steps in the design of various statistics.

A fully Bayesian treatment of the problem requires (and uniquely follows from) the explicit specification of competing models of an experiment, including an explicit statement of how plausible we find any particular combination of the model parameters.  This includes working out \emph{a priori} how to `spread our bets' over the large dimensional space of all possible signal waveforms.  Such \emph{prior plausibility distributions} are necessarily subjective---there is no `right' nor `wrong' way to choose them---but they do make definite statements about our expectations, and it is difficult to conceive of a scientist who would knowingly bet, for instance, that most detected gravitational waves will have a strain greater than unity.

The Bayesian requirement for a waveform prior stands in stark contrast to \gursel-Tinto's advertised independence of signal waveform.  The difference reflects a common criticism of Bayesian inference: that the use of priors introduces subjectivity and even bias into an analysis.  One response to this criticism is to note that non-Bayesian statistics are not free of prior expectations about the world, but that their
priors are merely implicit, unexamined and because of this may even contradict the stated intent of their designers.  This suggests the question: what choice of priors will make a Bayesian statistic behave like the \gursel-Tinto statistic and its relatives?

In this paper we set up a Bayesian analysis treating the experiment as a linear model with unspecified multivariate normal distributions as the priors on noise and signal.  We then demonstrate how the Bayesian statistic reduces to (or limits to) \gursel-Tinto and other related methods if we choose certain priors.  The results are astonishing: none of the methods assume a uniform signal population across the sky, and most assume gravitational waves are unphysically large or undetectably small and very frequent.  These priors are not `wrong', but they are very far from the best guess of the scientists that designed them.  Nor are the statistics necessarily ineffective, but they must be less effective than an practically implementable Bayesian statistic whose priors better reflect the real world.

\section{Bayesian formulation\label{sec:Bayesian}}
We can formulate much of the problem as a linear system.  Pack $N$ time series, each of an observatory's $M$ measurements, into a vector
\begin{equation}
\mathbf{x}=[x_{1,1},x_{1,2},\ldots,x_{1,M},x_{2,1},\ldots,x_{N,M}]^T.
\end{equation}
We could work equally well with any linear transformation of the time series (such as the Fourier or wavelet transforms).

One model for the noise in the system is that it is drawn from a multivariate normal distribution, described by an $MN\times MN$ covariance matrix $\mathbf{A}$.  The plausibility of $\mathbf{x}$ assuming the noise-only hypothesis is then
\begin{equation}
p(\mathbf{x}|\textrm{`noise'})=\frac{1}{(2\pi)^{MN/2}\sqrt{\det{\mathbf{A}}}}\exp-\frac{1}{2}\mathbf{x}^T\mathbf{A}^{-1}\mathbf{x}.
\end{equation}

We may search for a \emph{particular} signal
\begin{equation}
\mathbf{h}=[h_{+,1},h_{+,2},\ldots,h_{+,M},h_{\times,1},\ldots,h_{\times,M}]
\end{equation}
which produces an additive response $\mathbf{Fh}$, where the $MN\times2M$ response matrix $\mathbf{F}$ includes the antenna patterns and geometrical time delay associated with the \emph{particular} direction $(\theta,\phi)$ of the signal and the physical locations of the interferometers.
The distribution of the data expected in the presence of this signal is
\begin{equation}
p(\mathbf{x}|\mathbf{h})=\frac{1}{(2\pi)^{MN/2}\sqrt{\det{\mathbf{A}}}}\exp-\frac{1}{2}(\mathbf{x-Fh})^T\mathbf{A}^{-1}(\mathbf{x-Fh}).
\end{equation}
This form is of no direct use because we do not know any \emph{particular} signal $\mathbf{h}$ we want to search for.  Bayesian marginalisation solves this problem for us but requires us to place a prior plausibility distribution on $\mathbf{h}$:
\begin{equation}
p(\mathbf{x}|\textrm{`signal'})=\int_\mathbf{h}\mathrm{d}\mathbf{h}\,p(\mathbf{h})p(\mathbf{x}|\mathbf{h}).
\end{equation}
We must specify \emph{a priori} how likely we think any particular signal is to occur.  If the marginalisation integral is to be solved symbolically, this distribution must take the form of a multivariate normal distribution.  However, some desirable signal models (those where the signal has fewer than $2M$ degrees of freedom) will result in singular covariance matrices.  Instead we adopt a multivariate normal distribution with an $L\times L$ covariance matrix $\mathbf{Z}$ for abstract signal \emph{parameters} $\mathbf{y}$ related to $\mathbf{h}$ through a $2M\times L$ matrix $\mathbf{W}$ of \emph{template waveforms}, so that $\mathbf{h=Wy}$.

A source with known waveforms $\mathbf{w}_+$ and $\mathbf{w}_\times$ whose relative amplitudes and projections onto the detector's polarisation basis is unknown (as is the case for a source of unknown polarisation angle and inclination) could be modelled with four amplitude parameters, $\mathbf{Z}=\mathbf{I}$ and
\begin{equation}
\mathbf{W}=\left[
\begin{array}{cccc}
\mathbf{w}_+ & \mathbf{w}_\times & \mathbf{0} & \mathbf{0} \\
\mathbf{0} & \mathbf{0} & \mathbf{w}_+ & \mathbf{w}_\times
\end{array}
\right].
\end{equation}
This is a Bayesian analogue of a coherent matched filter analysis.

The \emph{least informative} multivariate normal distribution prior has $2M$ parameters (one for each sample of each polarisation), with $\mathbf{Z=I}$ and $\mathbf{W}=\sigma\mathbf{I}$ where $\sigma$ is a scale factor. This prior says that we know nothing about $h_i$ (save that it is unlikely to be very much larger that $\sigma$); we do not know if it is positive or negative or zero, and we do not know if it is correlated with $h_{j}$ positively or negatively or not at all.  This distribution also describes white noise, and we will call it a `white noise' signal prior, even though it detects all signals, not just those we would classify as `noise-like'.



With a signal prior defined we may now solve
\begin{eqnarray}
p(\mathbf{x}|\textrm{`signal'})&=&\int_\mathbf{y}\mathrm{d}\mathbf{y}\ p(\mathbf{y})p(\mathbf{x}|\mathbf{h})\\
&=&\int_\mathbf{y}\mathrm{d}\mathbf{y}\frac{\exp-\frac{1}{2}[(\mathbf{x-FWy})^T\mathbf{A}^{-1}(\mathbf{x-FWy})+\mathbf{y}^T\mathbf{Z}^{-1}\mathbf{y}]}{(2\pi)^{(MN+L)/2}\sqrt{\det{\mathbf{A}}\det{\mathbf{Z}}}}\nonumber\\
&=&\frac{1}{(2\pi)^{MN/2}\sqrt{\det{\mathbf{C}}}}\exp-\frac{1}{2}\mathbf{x}^T\mathbf{C}^{-1}\mathbf{x}\\
\textrm{where}\ \mathbf{C}^{-1}&=&\mathbf{A}^{-1}-(\mathbf{A}^{-1}\mathbf{FW})[(\mathbf{FW})^T\mathbf{A}^{-1}(\mathbf{FW})+\mathbf{Z}^{-1}]^{-1}(\mathbf{A}^{-1}\mathbf{FW})^T.\nonumber\label{eq:marginalizingh}
\end{eqnarray}
This is just another multivariate normal distribution.

The Bayesian odds ratio tells us how likely the `signal' hypothesis is, given the observation $\mathbf{x}$ in terms of how likely the observation is, given the two hypotheses under consideration and the prior plausibility of those hypotheses:
\begin{eqnarray}
O&=&\frac{p(\textrm{`signal'}|\mathbf{x})}{p(\textrm{`noise'}|\mathbf{x})}=\frac{p(\textrm{`signal'})}{p(\textrm{`noise'})}\frac{p(\mathbf{x}|\textrm{`signal'})}{p(\mathbf{x}|\textrm{`noise'})}\nonumber\\
&=&\frac{p(\textrm{`signal'})}{p(\textrm{`noise'})}\sqrt{\frac{\det{\mathbf{A}}}{\det{\mathbf{C}}}}\exp-\frac{1}{2}\mathbf{x}^T(\mathbf{C}^{-1}-\mathbf{A}^{-1})\mathbf{x}.
\end{eqnarray}
We should choose the prior odds ratio $p(\textrm{`signal'})/p(\textrm{`noise'})$ to be $\ll1$, reflecting the infrequency of detectable gravitational waves.  (This is analogous to choosing the threshold $\lambda$ for a frequentist likelihood.)

This is not yet a gravitational wave search; it looks for any gravitational wave of a known size from a known direction at a known time.  In \S\ref{sec:search} we will see how to perform a Bayesian search.  However, this odds ratio is the part of the Bayesian analysis that can be compared with previously proposed methods.

\section{Comparison with existing methods}
Several previously proposed tests 
have been expressed in the form
\begin{equation}
\mathbf{x}^T\mathbf{B}\,\mathbf{x}>\lambda
\end{equation}
for matrix $\mathbf{B}$ and threshold $\lambda$.  The Bayesian test is instead
\begin{eqnarray}
\frac{p(\textrm{`signal'})}{p(\textrm{`noise'})}\sqrt{\frac{\det{\mathbf{A}}}{\det{\mathbf{C}}}}\exp-\frac{1}{2}\mathbf{x}^T(\mathbf{C}^{-1}-\mathbf{A}^{-1})\mathbf{x}&>&1\label{eq:Bayesiantest}
\end{eqnarray}
or, rearranging Equation~\ref{eq:Bayesiantest},
\begin{eqnarray}
\mathbf{x}^T(\mathbf{A}^{-1}-\mathbf{C}^{-1})\mathbf{x}>-2\ln\frac{p(\textrm{`signal'})}{p(\textrm{`noise'})}\sqrt{\frac{\det{\mathbf{A}}}{\det{\mathbf{C}}}}\label{eq:lnOdds}
\end{eqnarray}
suggesting that by the appropriate choice of priors (including the signal model) we can create a Bayesian test that behaves like one of the previously proposed tests.

\subsection{Tikhonov regularised statistic}
Consider $\mathbf{A}=\mathbf{I}$, $\mathbf{W}=\sigma\mathbf{I}$ and $\mathbf{Z}=\mathbf{I}$ (white noise detectors and white noise signal of characteristic amplitude $\sigma$).  Also let
\begin{equation}
\frac{p(\textrm{`signal'})}{p(\textrm{`noise'})}=e^{-\lambda/2}\sqrt{\det(\sigma^{2}\mathbf{F}^T\mathbf{F}+\mathbf{I})}.
\end{equation}
As $\mathbf{F}$ varies with direction, so too does the prior.  For these priors, the Bayesian test of Equation~\ref{eq:lnOdds} becomes
\begin{equation}
\mathbf{x}^T\mathbf{F}(\mathbf{F}^T\mathbf{F}+\sigma^{-2}\mathbf{I})^{-1}\mathbf{F}^T\mathbf{x}
>\lambda\label{eq:tik}
\end{equation}
which is the Tikhonov regularised statistic of \cite{Ra:06} with regulariser $\alpha=\sigma^{-1}$.

The priors show that the Tikhonov regularised statistic implicitly assumes that gravitational wave bursts have a \emph{particular size} (this physical interpretation of the regulariser was not made in \cite{Ra:06}) and are \emph{distributed non-uniformly on the sky}. 
The strength of the directional bias depends on the size of the expected signal.

\subsection{\gursel-Tinto statistic}
If we take the limit $\sigma\rightarrow\infty$ (the large signal limit) Equation~\ref{eq:tik} limits to
\begin{equation}
\mathbf{x}^T\mathbf{F}(\mathbf{F}^T\mathbf{F})^{-1}\mathbf{F}^T\mathbf{x}
>\lambda
\end{equation}
which is the \gursel-Tinto statistic \cite{GuTi:89}.

The priors show that the \gursel-Tinto statistic implicitly assumes that gravitational wave bursts are \emph{very large},
\emph{very frequent} (note that $p(\textrm{`signal'})/p(\textrm{`noise'})\rightarrow\infty$), and \emph{distributed non-uniformly on the sky}.

The effect of these priors is not so dramatic as we might expect; \gursel-Tinto works in practice.  Signals of physically reasonable sizes are only an infinitesimal part of this signal population, but signals from this population occur infinitely frequently, and thus the method is able to detect them.  Even so, one of \gursel-Tinto's common failure modes is to misidentify a typical gravitational wave injection as a much larger (but, it believes, still likely to occur) gravitational wave with a different direction and polarisation that the network is nearly insensitive to.

\subsection{Soft constraint statistic}
Now allow the signal's characteristic amplitude to vary with direction, so that \mbox{$\mathbf{W}=\sigma k(\Omega)\mathbf{I}$}, and divide the Bayesian test in Equation~\ref{eq:lnOdds} through by $\sigma^2$:
\begin{eqnarray}
\frac{1}{\sigma^2}\mathbf{x}^T\mathbf{F}(\mathbf{F}^T\mathbf{F}+\frac{1}{\sigma^2 k^2(\Omega)}\mathbf{I})^{-1}\mathbf{F}^T\mathbf{x}> -\frac{2}{\sigma^2}\ln\frac{p(\textrm{`signal'})/p(\textrm{`noise'})}{\sqrt{\det[\sigma^2k^2(\Omega)\mathbf{F}^T\mathbf{F}+\mathbf{I}]}}\nonumber.
\end{eqnarray}
Let
\begin{equation}
\frac{p(\textrm{`signal'})}{p(\textrm{`noise'})}=e^{-\lambda\sigma^2/2}\sqrt{\det[\sigma^{2}k^2(\Omega)\mathbf{F}^T\mathbf{F}+\mathbf{I}]}.
\end{equation}
Then in the limit of $\sigma\rightarrow0$ we have
\begin{equation}
k^2(\Omega)\mathbf{x}^T\mathbf{FF}^T\mathbf{x}>\lambda
\end{equation}
which is the soft constraint statistic of \cite{KlMoRaMi:05} if the normalization function $k^{-2}(\Omega)$ is equal to the principal eigenvalue of $\mathbf{F}^T\mathbf{F}$ (though the normalization function $k^{-2}(\Omega)=\tr\mathbf{F}^T\mathbf{F}$ is required to give a direction-free null hypothesis distribution).

The priors show that the soft constraint statistic implicitly assumes gravitational wave bursts are \emph{very small}
and \emph{very frequent} (note that \mbox{$p(\textrm{`signal'})/p(\textrm{`noise'})\rightarrow 1$}).
The Bayesian statistic can only draw very weak conclusions about the presence of very small signals (the division of Equation~\ref{eq:lnOdds} by $\sigma^2$ is necessary to `blow up' its diminishing range), so even though the priors depend only weakly on direction they still strongly bias the soft constraint.

\subsection{Hard constraint statistic}
Now consider a linearly polarised signal model with a polarisation angle that is some known function of direction $\psi(\Omega)$ and $\mathbf{W}=\sigma k(\Omega)\mathbf{U}$ where
\begin{equation}
\mathbf{U}
=
\left[
\begin{array}{c}
\cos2\psi(\Omega)\,\mathbf{I}\\
\sin2\psi(\Omega)\,\mathbf{I}
\end{array}
\right].
\end{equation}
Let
\begin{equation}
\frac{p(\textrm{`signal'})}{p(\textrm{`noise'})}=e^{-\lambda\sigma^2/2}\sqrt{\det[\sigma^{2}k^2(\Omega)(\mathbf{FU})^T\mathbf{FU}+\mathbf{I}]}.
\end{equation}
Then in the limit of $\sigma\rightarrow0$ the Bayesian test Equation~\ref{eq:lnOdds} divided by $\sigma^2$ goes to
\begin{equation}
k^2(\Omega)\mathbf{x}^T\mathbf{FU}(\mathbf{FU})^T\mathbf{x}>\lambda.
\end{equation}
If the polarisation angle is defined as the one the network is most sensitive to, and the normalisation function $k^{-2}(\Omega)$ is taken to be the principal eigenvalue of $(\mathbf{FU})^T\mathbf{FU}$, this expression is the hard constraint statistic of \cite{KlMoRaMi:05}.

As well as the explicit design assumptions that gravitational wave bursts are \emph{linearly polarised} and \emph{optimally oriented}, the priors show that the hard constraint statistic implicitly assumes that gravitational wave bursts are \emph{very small}
 and \emph{very frequent} (note that $p(\textrm{`signal'})/p(\textrm{`noise'})\rightarrow 1$), and (like the soft constraint) is strongly affected by the weak directional dependency of its priors.

\subsection{Implications of implicit priors}
\begin{figure}
\resizebox{\columnwidth}{!}{\includegraphics{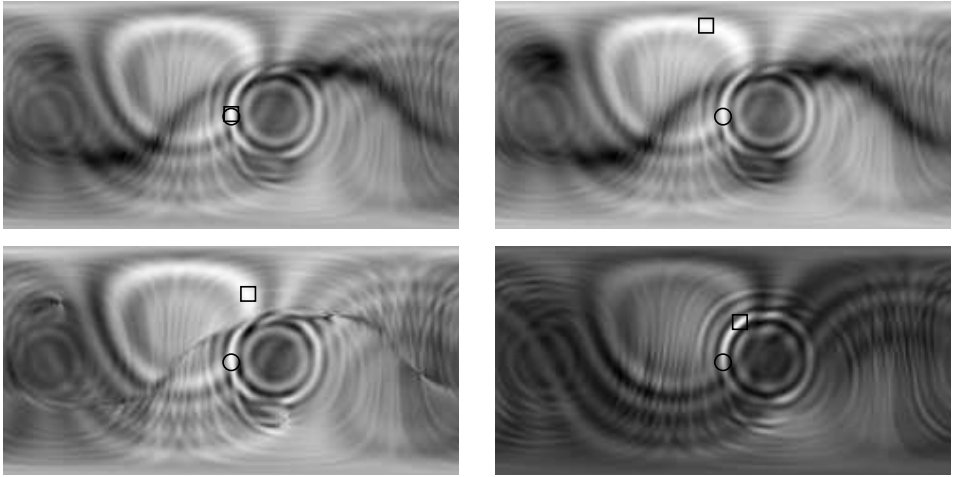}}
\caption{Four statistics as a function of direction $\Omega=(\theta,\phi)$ for an injection of 1/16 seconds of white noise with amplitude SNR 5 into 3 identical white noise detectors sampled at 1024 Hz with the locations and orientations of LIGO Hanford, LIGO Livingston and Virgo.  Black is least plausible; white is most plausible; a circle indicates the true direction; a square indicates the maximum of the statistic.  Top left: $\ln$ Bayesian odds ratio with $\sigma=5$; the signal is localised with 95\% confidence to $10^{-4}$ of the sky including the true direction.  Top right: Tikhonov with $\alpha=1/5$; this differs from the Bayesian case only by the direction prior.  Bottom left: \gursel-Tinto; note the discontinuities where the network becomes insensitive to one polarisation.  Bottom right: Soft constraint with $k(\Omega)^{-2}=\tr\mathbf{F}^T\mathbf{F}$.  For this injection, the global maximums of the non-Bayesian statistics are not consistent with the true direction.}\label{fig:skymaps}
\end{figure}
The unphysical priors required to make a Bayesian analysis behave like any of the previously proposed methods indicates that none of these methods are optimal for the realistic scenarios their designers were targeting; they are optimal for scenarios greatly different, where gravitational waves are infinitely large or small and not uniformly distributed on the sky.  A Bayesian analysis whose priors better reflect the true signal population will necessarily outperform all of these methods (see Figure~\ref{fig:skymaps}), but the analysis above does not indicate how much better it will do.

What went wrong in the design of these methods?  In the case of \gursel-Tinto, the good idea of waveform independence was translated into an uninformative improper prior; but if an observation can be explained by either a moderately sized signal or an unphysically enormous signal, it is not optimal to treat these as equally credible alternatives.
Ensuring that all directions have the same null hypothesis distribution counter-intuitively introduces a bias on the sky, because where the network is more sensitive the absence of excess power is actually stronger evidence against the signal hypothesis.  The `two detector paradox' described in \cite{KlMoRaMi:05} is simply an artefact of \gursel-Tinto's uninformative prior.  Of the \emph{ad hoc} attempts to fix the `paradox' described in \cite{KlMoRaMi:05,Ra:06}, only the Tikhonov regulariser constitutes an unambiguous theoretical improvement.

We have not considered here all the methods that have been proposed in the literature.  Some of these, notably the other regularisation method in \cite{Ra:06} and the almost-Tikhonov-regularisation in \cite{MoRaKlMi:06}, likely suffer from similar problems.

\section{Bayesian search}
\subsection{Robustness\label{sec:robustness}}

All the methods discussed so far (including the Bayesian test) consider the only source of excess energy in interferometers to be a gravitational wave.  Data orthogonal to $\textrm{span}\,\mathbf{F}$ is ignored because the noise and signal hypothesis make identical predictions and cannot be distinguished by it, even though that data is the \gursel-Tinto null stream \cite{GuTi:89} and excess energy in it rejects the gravitational wave hypothesis.  \cite{Ch_etal:06} supplemented \gursel-Tinto with an \emph{ad hoc} statistic in an attempt to solve this shortcoming.

A better approach (used in \cite{Cl_etal:07}) is to include interferometer `glitches' in the set of hypotheses under consideration; they are just another kind of signal model.  A glitch can be represented as a different noise covariance distribution replacing $\mathbf{A}$.  For example, if white noise glitches occur in white detectors with plausibility $p(\textrm{`glitch'})$, are additive and have characteristic amplitude $\gamma$, then we have a family of $2^N$ covariance distributions such as $\mathbf{A}_{01\ldots0}=\textrm{diag}[\mathbf{I},(1+\gamma^2)\mathbf{I},\ldots,\mathbf{I}]$ with prior plausibilities of the form $p(\textrm{`glitch'})^k[1-p(\textrm{`glitch'})]^{2^N-k}$.  The $N$-glitch hypothesis can `explain' any data, but it incurs a large \emph{Occam penalty} for its generality, whereas the signal hypothesis can more parsimoniously explain the tiny subspace of all possible data that is consistent with a gravitational wave.  Signals from directions (or with a polarisation) that one detector is insensitive to are forced to compete against the less penalised $N-1$ glitch hypotheses
If we have more knowledge about glitches, we can use a set of glitch waveforms and parameters analogously to the construction of our signal hypothesis.

\subsection{Marginalisation\label{sec:search}}

To perform a search over parameters the model is not linear in, like time-of-arrival $\tau$, sky direction $\Omega$ and signal model types, we need to numerically marginalise over these variables.  For example, we can produce the odds ratio for a signal from \emph{any} direction by marginalising over $\Omega$:
\begin{eqnarray}
O&=&\frac{p(\textrm{`signal'})}{p(\textrm{`noise'})}\frac{1}{p(\mathbf{x}|\textrm{`noise'})}\int_{S^2}\textrm{d}\Omega\ p(\Omega)p(\mathbf{x}|\textrm{`signal'}).
\end{eqnarray}
Previously proposed methods would instead \emph{maximise} their statistic over their parameters, so we cannot directly compare this step, other than to note that for confidently detected signals the marginalisation will be dominated by a single peak and the integral will be proportional to the maximum of the integrand.

Signals from sources distributed uniformly in space will have a power-law distribution of characteristic amplitudes $p(\sigma)\propto\sigma^{-4}$ (this encodes the `twice the sensitivity,
eight times the event rate' rule of thumb).  This provides us with a simple example of a parameterised set of signal models $\mathbf{W}=\sigma\mathbf{W}_1$ and a physically-inspired prior on the parameter.  We may also want to look for signals parameterised in other ways.  For example, we may want to consider different time and frequency bounds on time- and/or frequency-band-limited signals.

\subsection{Implementation}
If the noise and signal models are stationary, the Bayesian statistic for a particular signal size can be implemented (using a transformation to the Fourier domain that diagonalises $\mathbf{C}$) as efficiently as the coloured noise versions of the previously proposed statistics considered here.  If we declare the glitch and signal hypotheses exclusive (so that we will reject the very infrequent gravitational wave that occurs at the same time as a glitch) and model glitches in each detector as independent, we incur only a minimal extra cost to gain robustness against that population of instrumental artefacts.  Marginalising over $\sigma$ can reuse many intermediate values and requires relatively few samples for a steep $\sigma^{-4}$ prior.  Many other signal models present comparable opportunities for optimisation.

\section{Conclusion}
A simple Bayesian analysis of the problem of coherent detection of unmodelled gravitational wave bursts with a network of ground-based interferometric gravitational wave detectors is presented.  It reduces or limits to several previously proposed methods for the same problem only when unphysical priors are used, indicating that those previously proposed methods cannot be optimal for realistic signal populations.  A method for improving the robustness of the Bayesian method by substituting a more realistic noise model was noted, as were the steps for generalising to an all-sky unknown-arrival-time search.  The Bayesian method can be implemented, depending on the signal model, to be computationally competitive with existing methods.

We can contrast the process of specifying physical models of the world, quantifying their predictions and forming the Bayesian statistic in \S\ref{sec:Bayesian} with the looser process of heuristically arguing our way to a statistic \cite{GuTi:89} and then noting its flaws and heuristically arguing our way to a mutated statistic \cite{Ch_etal:06,KlMoRaMi:05,Ra:06}.  In both cases, designers have freedom and subjective choices are made.  In the Bayesian case, these are restricted to choosing the physical model and its priors; these have immediate physical interpretations which are explicit and can be contested.  The Bayesian statistic follows uniquely from them.  In the non-Bayesian case choices are made at every stage and their implications are uncertain,
leaving us open to unexamined, unintended and unfortunate consequences.

\section*{Acknowledgments}
We would like to thank 
Shourov Chatterji, Albert Lazzarini
and Andrew Moylan.
Antony Searle and Graham Woan were supported by the LIGO Visitors Program.  Antony Searle was supported by the Australian Research Council.  Part of this research was performed at the Jet Propulsion Laboratory, California Institute of Technology, under contract with the National Aeronautics and Space Administration. Massimo Tinto was supported under research task 05-BEFS05-0014.  Patrick J Sutton  was supported in part by STFC grant PP/F001096/1.  LIGO was constructed by the California Institute of Technology and Massachusetts Institute of Technology with funding from the National Science Foundation and operates under cooperative agreement PHY-0107417.  This document has been assigned LIGO Document Number LIGO-P070119-00-Z.
\section*{References}
\bibliography{main}

\begin{thebibliography}{1}

\bibitem{GuTi:89}
Y.~Gursel and M.~Tinto.
\newblock {\em Phys. Rev. D}, 40:3884, 1989.

\bibitem{Ch_etal:06}
S.~Chatterji, A.~Lazzarini, L.~Stein, P.~Sutton, A.~Searle, and M.~Tinto.
\newblock Coherent network analysis technique for discriminating
  gravitational-wave bursts from instrumental noise.
\newblock {\em Phys. Rev. D}, 74:082005, 2006.

\bibitem{KlMoRaMi:05}
S.~Klimenko, S.~Mohanty, M.~Rakhmanov, and G.~Mitselmakher.
\newblock {\em Phys. Rev. D}, 72:122002, 2005.

\bibitem{KlMoRaMi:06}
S.~Klimenko, S.~Mohanty, M.~Rakhmanov, and G.~Mitselmakher.
\newblock Constraint likelihood method: Generalization for colored noise.
\newblock {\em J. Phys. Conf. Ser.}, 32:12, 2006.

\bibitem{MoRaKlMi:06}
S.~Mohanty, M.~Rakhmanov, S.~Klimenko, and G.~Mitselmakher.
\newblock Variability of signal to noise ratio and the network analysis of
  gravitational wave burst signals.
\newblock {\em Class. Quant. Grav.}, 23:4799--4809, 2006.

\bibitem{Ra:06}
M.~Rakhmanov.
\newblock Rank deficiency and tikhonov regularization in the inverse problem
  for gravitational-wave bursts.
\newblock {\em Class. Quant. Grav.}, 23:S673--S685, 2006.

\bibitem{AnBrCrFl:01}
W.~G. Anderson, P.~R. Brady, J.~D.~E. Creighton, and \'E.~\'E. Flanagan.
\newblock {\em Phys. Rev. D}, 63:042003, 2001.

\bibitem{Cl_etal:07}
J.~Clark, Ik~Siong Heng, M.~Pitkin, and G.~Woan.
\newblock {\em Phys. Rev. D}, 76:043003, 2007.

\end{thebibliography}
\end{document}